# Size and Shape of a Celestial Body – Definition of a Planet

Konrad Probsthain

Johannes-Lampe-Str. 9, D 64823 Gross-Umstadt, Germany

E-mail address: kplaplace@aol.com

**Abstract**

Experience shows that celestial bodies have a 'nearly round' shape only from a certain size. At IAU Resolution B5, item (b), that shape serves as an indicator for a distinct mechanism of its forming, caused by a minimum of mass. Rigid-body forces should have been overcome by self-gravity and a hydrostatic equilibrium shape should have been achieved.

A new approach to correlate the size and shape of a solid in hydrostatic equilibrium by balancing self-gravity and rigid-body forces leads to a real, not to an arbitrary lower limit of size. No arbitrary criterion is required as it is often the case. Above this limit the shape of a solid body is restricted by a maximum of its surface area, and this maximum vanishes only at infinite size. Therefore the shape like that of a related fluid in hydrostatic equilibrium can only be reached in the solid state, if its surface area is larger than this maximum. This applies only to the four giant planets and Haumea. All the known 'nearly round' celestial bodies in the solar system (the dwarf Haumea is a questionable exception) can't have achieved their present shape while being solids. The present shapes have been formed before solidification and are frozen. The gas giants are not solids at all and the terrestrial planets have at least partially melted in their thermal history for millions of years. The smaller 'nearly round' objects such as asteroids and satellites were at least partially melted combined with internal differentiation and resurfacing of a mechanically unstable crust in their very early thermal history. No rigid-body forces had to be overcome.

Item (b) of the current planet definition is somewhat undetermined. All planets and dwarfs in the solar system match this requirement only at very generous interpretation. It should be completely deleted. The number of dwarfs could be restricted by an arbitrary minimum of mass.

## 1. Introduction

The IAU Resolution B5 of 2006 defines planets in the solar system on the basis of the following conditions:

(1) A planet is a celestial body that
    (a) is in orbit around the Sun,
    (b) has sufficient mass for its self-gravity to overcome rigid body forces so that it assumes a hydrostatic equilibrium (nearly round) shape, and
    (c) has cleared the neighbourhood around its orbit.
(2) A 'dwarf planet' is a celestial body that
    (a) is in orbit around the Sun,
    (b) has sufficient mass for its self-gravity to overcome rigid body forces so that it assumes a hydrostatic equilibrium (nearly round) shape, and
    (c) has not cleared the neighbourhood around its orbit, and
    (d) is not a satellite.

By the condition (c), the distinction between planets and so-called dwarf planets is delineated. This has led to violent discussions and emotional criticism, because Pluto can't be a planet anymore (WEINTRAUB 2008).

The condition (b), however, has so far hardly been questioned. Only SARMA et al. (2008) criticize this point as problematic, confusing and non-essential. They recommend its complete removal from the planet definition. On the other hand BROWN (2008) insists on a nearly round shape as a suitable criterion for distinction between dwarf planets and other celestial objects.

Experience has shown that celestial bodies look approximately spherical with an average radius of at least 200 km (ice moons) or 300 km (asteroids).

There were only some efforts to quantitatively link the four main features of mass, self-gravitation, rigid-body forces, and hydrostatic equilibrium with the derived feature of the spherical shape. TANCREDI & FAVRE (2008) have summarized previous work (JOHNSON & MCGETCHIN 1973, FARINELLA et al. 1983, COLE 1984, SLYUTA & VOROPAEV 1997) and found general accordance except small quantitative deviations. LINEWEATHER & NORMAN 2010 and O'CALLAGHAN 2012 come to similar results. A certain limit for a more irregular to a more spherical shape transition is predicted, although the criterion for such a transition borderline is always arbitrary. But even the IAU admits in a press release that "all borderline cases would have to be established by observation".

JOHNSON & MCGETCHIN 1973 consider a stable surface roughness at hydrostatic equilibrium of a solid body. The other authors choose a 'tolerable' deviation of the main dimensions of the body from its mean size as a criterion for the potato-to-sphere transition. The limit values determined in this way are compatible with the observation, but that is not a real limit.

Departures from hydrostatic equilibrium shape of a liquid body, i. e. sphere or oblate, are also considered by THOMAS et al. (2007), because they may reveal information on processes that complicate or defeat tendencies to relax to equilibrium shapes. The present form of celestial bodies (here Saturnian icy satellites) could be approached at any point in their evolution. Rigid body forces to be overcome are not explicitly considered.

LINEWEATHER & NORMAN (2010) like their predecessors apply the yield strength of the relevant material, rock or ice, for the rigid body forces to be overcome. (SCHEUER (1981) chose a similar approach to estimate the maximum height of mountains.). But Holsapple (2004) ignores tensile strength at a cohesionless model to equilibrium figures of spinning bodies with self-gravity.

In deed the wording of the condition (b) is somewhat complicated and not concrete. Obviously there should be a lower limit for mass and size, but without giving a real number. Instead of a concrete limit the mass is associated with self-gravity, rigid-body forces and hydrostatic equilibrium that should result in a 'nearly round' shape. The latter feature is the only one that could be confirmed by observation, if at all. But even if a body has an almost round shape, it is not certain whether rigid-body forces have been overcome by self-gravity and a hydrostatic equilibrium has been achieved.

There remains a certain scope for the interpretation of the condition (b) besides the definition of what is 'nearly round':

(b1)    The celestial body has really achieved a 'nearly round' hydrostatic equilibrium shape by overcoming of rigid-body forces, i. e. in solid state.

(b2)    The celestial body is able to achieve a 'nearly round' hydrostatic equilibrium shape by overcoming of rigid-body forces, but that shape has been achieved before solidification.

(b3)    The celestial body has really achieved a 'nearly round' hydrostatic equilibrium shape, but it is not able to assume that shape by overcoming of rigid-body forces.

However, the hydrostatic equilibrium does not occur spontaneously in a viscous medium, it takes time. This point is often neglected. Only JOHNSON & MCGETCHIN (1973) have shown that small rocky bodies need a relaxation time of the same order of magnitude as the age of the Solar System.

## 2. Linking size and shape of solids - a new approach

### 2.1. Hydrostatic equilibrium in a solid body

Hydrostatic equilibrium is originally the final state of an ideal liquid apart of extern forces. It results in a perfectly spherical shape. The shape can be altered by rotation, inhomogeneity and/or the gravitational attraction of other bodies. Then the resulting stress in hydrostatic equilibrium must be zero.

$$\sigma_0 = \sigma_{sph} - \sigma_{rot} - \sigma_{inhom} - \sigma_{gr} - \cdots = 0 \qquad (1)$$

σ Stress    (Subscripts: 0 hydrostatic equilibrium, sph sphere, rot rotation, inhom inhomogeneity, gr gravitational attraction)

The equilibrium shape of a rotating and orbiting fluid body is generally that of a triaxial ellipsoid. Any shape with a surface area different of $A_{equ}$ is unstable. That behavior is often expected also for solid bodies as planets, asteroids and satellites, provided they have mass enough to overcome rigid-body forces (THOMAS et al. 2007, TANCREDI & FAVRE 2008, TRICARICO 2014).

Off course, solid bodies can also approach a hydrostatic equilibrium, because they behave as viscous mass, as long as the gravitational forces exceed the rigid-body forces. In that case the hydrostatic equilibrium shape achieved may differ from that of a related fluid.

For simplicity here we deal only with a homogeneous body free of external forces and not rotating.

$$\sigma_{eff} \geq \sigma_{sph} - \sigma_{pot} - \sigma_{rbf} \qquad (2)$$

$$\sigma_{eff} \geq \frac{Mg_{sph}}{A_{sph}} - \frac{Mg_{pot}}{A_{pot}} - \sigma_{rbf}$$

$$\sigma_{eff} \geq \frac{(\frac{4}{3}\pi R_{sph}^2 \rho)^2 G}{A_{sph}} - \frac{(\frac{4}{3}\pi R_{sph}^2 \rho)^2 G}{A_{pot}} - \sigma_{rbf} \qquad (3)$$

with $M = \frac{4}{3}\pi R_{sph}^3 \rho$ and $g_{sph} = \frac{4}{3}\pi R_{sph}\rho G$.

Mass: $M = M_{pot} = M_{sph}$, mean surface gravity: $g = g_{sph} \approx g_{pot}$, density: $\rho = \rho_{pot} = \rho_{sph}$, stress: $\sigma_{pot} < \sigma_{sph}$ (surface load), $\sigma_{rbf} = const$ (yield strength) R radius:, surface area: $A_{pot} > A_{sph}$, G gravitational constant (Subscripts: pot potato, sph sphere, rbf rigid-body forces, eff effective, 0 equilibrium)

$$\sigma_0 = \sigma_{eff} = 0 \qquad (4)$$

$$\sigma_0 = \pi(\frac{2R_{sph}\rho}{3})^2 G (1 - \frac{1}{(1+A_0/A_{sph})}) - \sigma_{rbf} \qquad (5)$$

with $A_0 = A_{pot} - A_{sph}$ at $\sigma_0$

At equilibrium solid bodies sustain a shape more or less different of a perfect sphere. That deviation can be characterized by the relation:

$$A_{pot}/A_{sph} - 1 \geq A_0/A_{sph} \qquad (6)$$

$$A_0/A_{sph} = 1/(R_{sph}^2 \rho^2 \frac{4\pi G}{9\sigma_{rbf.}} - 1) > 0 \qquad (7)$$

$$R_{sph} = \frac{3}{2\rho}\sqrt{\frac{\sigma_{rbf}}{\pi G}}\sqrt{\frac{1}{A_0/A_{sph}} + 1} \qquad (8)$$

If the surface $A_{pot}$ of a solid having an equal volume as a related sphere does not exceed the relationship (6), each shape is stable. A 'potato' consisting of two spherical calottes (the smaller one as a half sphere) named as a 'sphere with a bump' may illustrate the deviation in shape. For that case LINEWEAVER & NORMAN (2010) derived a similar formula for the 'potato'-radius as (8), where $B$ is a pre-factor and $f$ is a 'sphericity' parameter:

$$R_{pot} \approx B\left(\frac{\sigma_y}{G\rho^2}\right)^{\frac{1}{2}} \quad \text{with} \quad B = \sqrt{\frac{3}{2\pi(f^2-1)}} \qquad (9)$$

$f$ is taken as the ratio of the longest (half?) axis of the 'potato' to its average radius and relates to $A_0/A_{sph}$ for the same geometric body, i. e. that 'sphere with a bump'. But there is a big difference between formula (8) and (9) and others.

The results of all earlier attempts to find a critical radius can be summarized in a common formula, as already stated by TANCREDI & FAVRE (2008). To include (8) and (9), this common formula can be rearranged, where $\alpha$ is a slightly varying pre-factor and $F(f)$ is a shape function of the 'sphericity' parameter $f$.

$$R_{sph} = \alpha/\rho \sqrt{\frac{3\sigma_{rbf}}{\pi G}} \; F(f) \qquad (10)$$

$f$ depends on the geometry of the model used. At all previous work this parameter characterizes only a small linear deviation from the radius of a sphere. Therefore, setting a lower limit is always arbitrary, e. g. LINEWEAVER & NORMAN (2010) set $f$ = 1.1 in formula (9).

Unlike formula (9) and others formula (8) allows the specification of a finite lower limit for the hydrostatic equilibrium in a solid:

$$\lim_{f=A_0/A_{sph}\to\infty}(R_{sph}) = \frac{3}{2\rho}\sqrt{\frac{\sigma_{rbf.}}{\pi G}} \qquad (11)$$

Choosing the yield strength of rocks and ice with 10 and 5 MPa respectively as $\sigma_{rbf}$ and mean densities of 3.54 kg dm$^{-3}$ for rocky asteroids and 1.15 kg dm$^{-3}$ for icy moons in formula

(11) the minimum mass of such a solid body in hydrostatic equilibrium is 11.93 * $10^{18}$ kg or 39.1 * $10^{18}$ kg respectively. These figures correspond with sphere radii of 93 km and 201 km. They can only serve as the order of magnitude because of the uncertainty of $\sigma_{rbf}$. Certainly, the yield strength is greater than 1 MPa (TANCREDI & FAVRE 2008)..

Below these radii any shape would be stable, a sponge just like a teacup or a bowling ball, because it remains no overburden gravitational stress. There is no other definite boundary line above these values. But the deviation from a spherical shape is rapidly limited with increasing radius and tends to zero (Fig. 1). In other words, the condition for a stringent spherical shape is an infinite radius.

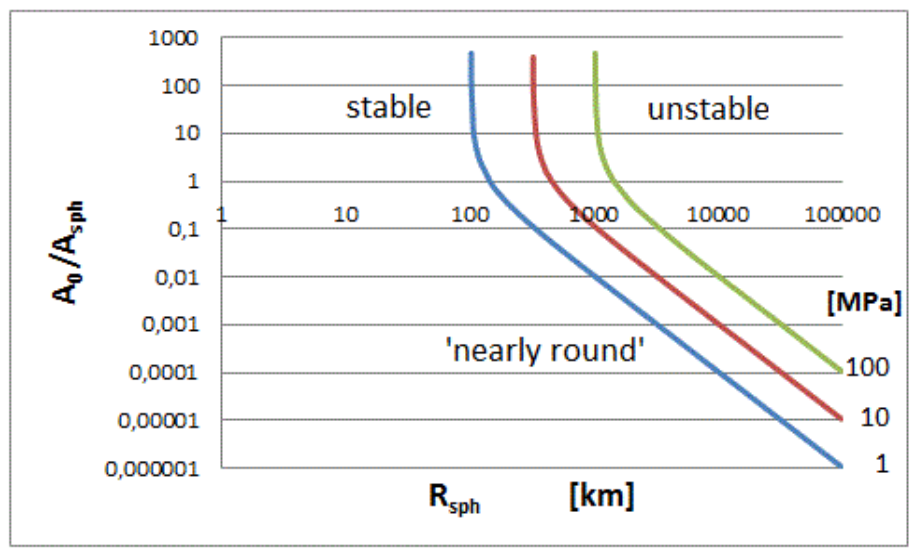

Fig. 1: Boundary lines for a hydrostatic equilibrium shape of a homogeneous solid when overcoming rigid-body forces by self-gravity, characterized by the surface area quotient $A_0/A_{sph}$ according formula (7). Parameter yield strength, abscissa mean radius at density 1 g/cm³.

The hydrostatic equilibrium shape of a related fluid body is that of an ellipsoid in its general sense. The surface area of such an ellipsoid can be larger or smaller than the limited surface area of the solid.

By overcoming rigid forces, a body in the solid state can only reach the hydrostatic equilibrium shape of the associated fluid if the surface of this ellipsoid is larger than this limit.

$$A_{pot} \geq A_{sph} + A_0 \qquad (12)$$

Otherwise, any shape of solid body below this surface boundary is stable and may include the hydrostatic equilibrium shape of a related fluid body.

But it does not have to look "nearly round". Where is the boundary?

Obviously the platonic solids, tetrahedron, cube etc., are very different from sphere in shape. Also a 'potato' as defined above doesn't look like a sphere, if $A_{pot}/A_{sph} - 1$ is near its maximum value 0.02599 for a 'sphere with a bump' (small bump = big bump). Otherwise a truncated icosahedron (soccer ball) or a sphere with many dimples (golf ball) is certainly accepted as round, despite the 'potato numbers' $A_{pot}/A_{sph} - 1$ are of the same range (0.03453, 0.00811) (Table 1).

| Body | Faces | Surface area | Mid sphere | 'Potato number' | Nearly round? (mutual) |
|---|---|---|---|---|---|
| | n | $A_{polyh}$ | $R_m$ | $A_{pot}/A_{sph}$-1 | |
| Tetrahedron | 4 | 155.24 | - | 0.4900 | No |
| Cube | 6 | 129.266 | 3.282 | 0.2407 | No |
| Octahedron | 8 | 123.214 | 2.982 | 0.1826 | No |
| Dodecahedron | 12 | 114.435 | 3.082 | 0.09835 | No |
| Icosahedron | 20 | 110.918 | 2.895 | 0.06459 | ? |
| Icosidodecahedron | 32 | 109.553 | 2.975 | 0.05149 | Yes ? |
| Truncated Icosah.* | 32 | 107.786 | 2.957 | 0.03453 | Yes |
| Hexakisoctahedron | 48 | 107.513 | 2.858 | 0.03191 | Yes |
| Pentakisdodecahe. | 60 | 106.37 | 2.88 | 0.02094 | Yes |
| Sphere with dimples ** | | | | 0.00811 | Yes |

Volume $V_{polyh}=V_{sph}$, $V_{sph}=100$, $A_{sph}=104.188$, $R_{sph}=2.879$
  * Soccer ball, n=32, 12 regular pentagons, 20 regular hexagons
  ** Golf ball, typical example 42.6 mm Ø, 450 dimples 3.0 mm Ø, 0.25 mm depth

Table 1: Centrally symmetric bodies, nearly round or not?

The decisions of the IAU, whether or not an object is accepted as dwarf planet, are probably not depending on a 'nearly round' shape alone but also that the shape is consistent with a fluid in hydrostatic equilibrium (not required in condition (b) of the planet definition!). Ceres at a 'potato number' of 0.001 is accepted as a dwarf planet, Pallas and Vesta at 0.003 are not, while Haumea at 0.065 is accepted, though looks more like a deformed disc or a fat cigar than a ball.

In every case the acceptance of a body as 'nearly round' is more a question of esthetics than of physics. Is a smooth surface without corners and edges sufficient? Is the similarity with a sphere deciding? Is a golf ball nearly round and an egg is not? Or are both acceptable?

Besides the static aspect of the equilibrium shape, the time scale of its relaxation is of interest.

### 2.2. Adaptation of the hydrostatic equilibrium in a viscous mass

Rigid body forces and hydrostatic equilibrium assume as contradictions by itself, but sub solidus convection is accepted in the solid Earth because it behaves as a viscous mass. Nevertheless the adaptation of the hydrostatic equilibrium needs time, the higher the viscosity and the smaller the mass (self-gravity) of the body.

$$\frac{d(\sigma_{eff}/\sigma_0 - 1)}{dt} = \frac{d(A_{pot}/(A_{sph}+A_0) - 1)}{dt} \sim \frac{\sigma_{eff} - \sigma_0}{\eta} =$$

$$\frac{\left(\frac{4}{3}\pi R_{sph}^2 \rho\right)^2 G \left(\frac{1}{A_{sph}+A_0} - \frac{1}{A_{pot}}\right)}{\eta} = \frac{\left(\frac{2}{3}R_{sph}\rho\right)^2 \pi G \left(1 - \frac{A_{sph}+A_0}{A_{pot}}\right)}{(1+A_0/A_{sph})\eta} \quad (13)$$

$$dt = \frac{\eta(1+A_0/A_{sph})}{\pi G \left(\frac{2}{3}R_{sph}\rho\right)^2} \left(1 + \frac{1}{A_{pot}/(A_{sph}+A_0)-1}\right) d(A_{pot}/(A_{sph}+A_0)-1) \quad (14)$$

$$\tau = \frac{9}{4}\frac{\eta}{\pi G R_{sph}^2 \rho^2}\left(1 + \frac{A_0}{A_{sph}}\right)\int_{\xi=1}^{e}\left(1 + \frac{1}{\xi}\right)d\xi \text{ with } \xi = \frac{A_{pot}}{A_{sph}+A_0} - 1 \quad (15)$$

$$\tau = \frac{9}{4}e\frac{\eta}{\pi G R_{sph}^2 \rho^2}\left(1 + \frac{A_0}{A_{sph}}\right), \quad \frac{9}{4}e \cong 6.12 \quad (16)$$

$\tau$ relaxation time, $\eta$ dynamic viscosity, $\nu = \eta/\rho$ kinematic viscosity

> (For comparison: JOHNSON & MCGETCHIN (1973) set for the relaxation time of 'a non-equilibrium equatorial bulge': $\tau = \frac{19\nu}{2gR_m} \cong 7{,}13\frac{\eta}{\pi G R_{sph}^2 \rho^2}$ ; $R_m$ mean Radius)

That simple relationship yields a time scale of only $10^4$ years for an Earth like body and nearly $10^7$ years for a rocky 'dwarf planet' of 300 km radius. In every case, $\tau$ is very short compared with observable tectonic time scales on the Earth, that are $10^7 - 10^8$ years (age of oceanic crust, velocity at the trench 2 – 15 cm/yr (MORRA et al. 2016)).

The crucial parameter calculating $\tau$ is the viscosity. In the Earth's mantle the viscosity is $\eta \geq 10^{20}$ Pas (STEINBERGER & CALDERWOOD 2006) and the viscosity of ice near its melting temperature is only $\eta \approx 10^{10}$ Pas (SOTIN & POITIER 1987). Therefore an icy object should immediately get its equilibrium shape. But at lower temperature the viscosity can reach $10^{28}$ Pas even for ice (JOHNSON & McGETCHIN 1973). Therefore small icy trans-Neptunian objects (TNO) would also not relax in a geological time scale.

Note that the formula (16) applies only to $\sigma_{eff} > \sigma_0$. If the shape of a solid body is initially "rounder" than in the hydrostatic equilibrium, nothing happens. The initial shape is 'frozen'. Later alteration of the initial shape by tectonics, volcanism or impacts can remain stable in a wide range.

With only one exception all planets and dwarf planets, classified or candidates, could not have been formed by overcoming rigid-body forces. Either there were no rigid-body forces in their early history, and they maintain a "frozen" form or are not solid at all (Table 2, Fig. 2).

During accretion, a spherical shape becomes more likely due to random effects. Solid particles assembled by adhesion can arrange a 'nearly round' rubble pile (COMITO et al. 2011). Not least, the present solid body may have been liquid in its thermal history.

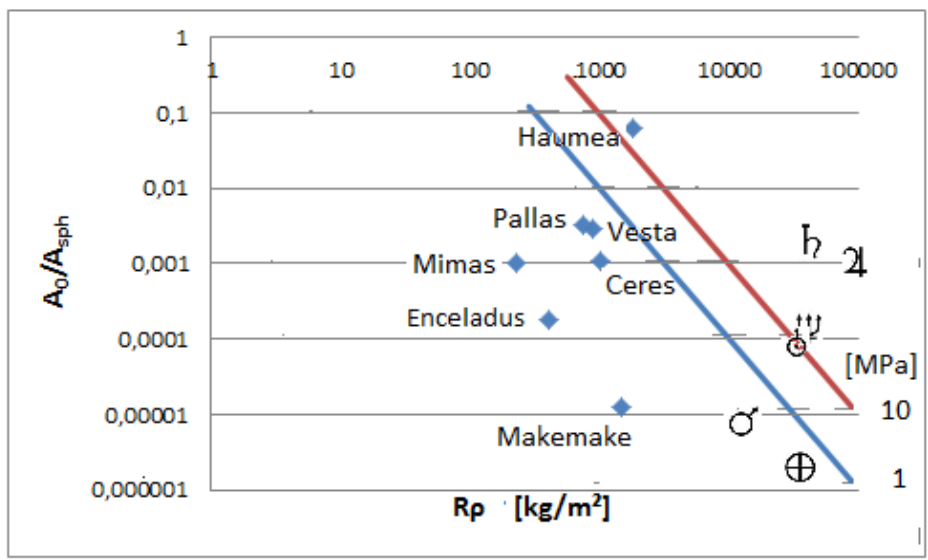

Fig. 2: Present shape of some 'nearly round' celestial bodies, characterized by the surface area quotient $A_{ellipsoid}/A_{sph} -1 >< A_0/A_{sph}$. Parameter of the borderlines: 1 and 10 MPa, density 1 g/cm³. Abscissa: mean radius times mean density.

| Celestial body | Mean radius (sphere) / Semi axis (ref. ellipsoid) [km] | Density [kg/m³] | Yield strength (estimate) [MPa] | $A_0/A_{sph}$ / $A_{ellips.}/A_{sph} - 1$ | Planet definition Case |
|---|---|---|---|---|---|
| Jupiter | 69,925 <br> Equ. 71,492 Pol. 66,850 | 1,326 <br> - | << 1 <br> - | << 0.125 $10^{-5}$ <br> 81.48 $10^{-5}$ | (b2) <br> Planet |
| Saturn | 58,240 <br> Equ. 60,268 Pol. 53,906 | 0,687 <br> - | << 1 <br> - | << 0.067 $10^{-4}$ <br> 22.68 $10^{-4}$ | (b2) <br> Planet |
| Uranus | 25,367 <br> Equ. 25,559 Pol. 24,975 | 1,270 <br> - | << 1 <br> - | << 0.103 $10^{-4}$ <br> 0.960 $10^{-4}$ | (b2) <br> Planet |
| Neptune | 24,623 <br> Equ. 24,764 Pol. 24,102 | 1,638 <br> - | << 1 <br> - | << 0.060 $10^{-4}$ <br> 1.320 $10^{-4}$ | (b2) <br> Planet |
| Earth | 6,371.01 <br> Equ. 6,378 Pol. 6,356.8 | 5,515 <br> - | 10 <br> - | 8.674 $10^{-5}$ <br> 0.198 $10^{-5}$ | (b2)? (b3) <br> Planet |
| Venus | 6,052 <br> No flattening | 5,243 <br> - | 10 <br> - | 9.633 $10^{-5}$ <br> << $10^{-5}$ | (b3) <br> Planet |
| Mars | 3,389 <br> Equ. 3,395 Pol. 3,378 [1]) | 3,930 <br> - | 10 <br> - | 6.025 $10^{-4}$ <br> 0.0447 $10^{-4}$ | (b3) <br> Planet |
| Mercury | 2,439.301 <br> 2,439.0+422x304x178 [2]) | 5,430 <br> - | 10 <br> - | 6.117 $10^{-4}$ <br> << $10^{-4}$ | (b3) <br> Planet |
| Moon | 1,737.4 <br> Equ. 1,738 Pol. 1,736 | 3,341 <br> - | 10 <br> - | 31.93 $10^{-4}$ <br> 0.0026 $10^{-4}$ | (b3) <br> Satellite |
| Eris | 1,163 <br> - | 2,25 <br> - | < 10 <br> - | < 1.582 $10^{-2}$ <br> << $10^{-2}$ | (b3) <br> Dwarf planet |
| Pluto | 1,187 <br> - | 1,86 <br> - | < 10 <br> - | < 2.251 $10^{-2}$ <br> << $10^{-2}$ | (b3) <br> Dwarf planet |
| Makemake | 715.0 <br> Equ. 717 Pol. 711 [3]) | 2,100 <br> - | < 10 <br> - | < 4.998 $10^{-2}$ <br> 0.0013 $10^{-2}$ | (b3) <br> Dwarf planet |
| Haumea | 715.2 <br> 960x770x495 [4]) | 2,600 <br> - | < 10 <br> - | < 3.203 $10^{-2}$ <br> 6.473 $10^{-2}$ | (b1)? (b2) <br> Dwarf planet |
| Ceres | 469.7 <br> 483x481x446 | 2,161 <br> - | < 10 <br> - | < 11.63 $10^{-2}$ <br> 0.109 $10^{-2}$ | (b3) <br> Dwarf planet |
| Pallas | 272.5 <br> 291x278x250 | 2,761 <br> - | < 10 <br> - | < 2.34 $10^{-1}$ <br> 0.033 $10^{-1}$ | (b3) <br> Asteroid |
| Vesta | 258.6 <br> 280x272x227 | 3,456 <br> - | 10 <br> - | 1.552 $10^{-1}$ <br> 0.029 $10^{-1}$ | (b3) <br> Asteroid |
| Enceladus | 252.1 <br> 257x251x248 [5]) | 1,609 <br> - | 5 <br> - | 4.839 $10^{-1}$ <br> 0.0018 $10^{-1}$ | (b3) <br> Satellite |
| Mimas | 198.2 <br> 208x197x191 [5]) | 1,149 <br> - | 5 <br> - | >> $10^{-1}$ <br> 0.01 $10^{-1}$ | (b3) <br> Satellite |

[1]) ARDALAN et al. 2010, [2]) KARIMI et al. 2015,
[3]) BROWN 2013, [4]) LOCKWOOD et al. 2014, [5]) THOMAS 2010

Table 2: Planets, dwarf planets and some ‚near-round' asteroids and satellites in relation to the requirement (b) of the planet definition.
Current surface area larger or smaller than the limit in overcoming rigid-body forces?

## 2.3. Mass and thermal history of a celestial body

The following heat flux balance is valid for the entire thermal history of a celestial body:

$$\frac{dT}{dt} M c = - \dot{Q}_S + \dot{Q}_G + \dot{Q}_R + \dot{Q}_L + \dot{Q}_C + \dot{Q}_I + \dot{Q}_T + \dots \qquad (17)$$

(Subscripts: S surface, G gravitational, R radiogenic, L latent, C core, I impact, T tidal...)

Equation (17) is the basis for each parametrized modeling of the thermal history of celestial bodies and has proved particularly useful in the modeling of sub-solidus convection in the terrestrial planets (e. g. SCHUBERT et al. 2001).

Some of the above terms, related to the heat capacity of the body, depend on its surface area, its rate of acceleration, its differentiation in core and mantle, but others do not. For a spherical celestial body with the radius $R_P$ and the core radius $R_C$, the following correlations apply:

$$\dot{Q}_S/(M\,c) \sim R_P^{-1}, \quad \dot{Q}_G/(M\,c) \sim R_P^{-1}\frac{dR_P}{dt}, \quad \dot{Q}_C/(M\,c) \sim R_C^2 R_P^{-3}\frac{dR_C}{dt},$$

$$\dot{Q}_R/(M\,c) \not\sim R_{P,C}, \quad \dot{Q}_L/(M\,c) \not\sim R_{P,C}, \quad \dot{Q}_I/(M\,c) \not\sim R_{P,C}, \dots \qquad (18)$$

There is a competition between the surface heat flux and the inner heat sources as accretion, segregation, radioactivity, impaction etc. The higher the accretion rate and the bigger the final mass of the body the higher is the peak temperature of its thermal history.

For the formation of a homogeneous, spherical planet of the mean density $\rho$ and the radius $R$ is (HEINTZ 2011):

$$\Delta T_{max} = \frac{Q_G}{M_R C_v} = G\,\frac{4\,\pi}{5}\,\frac{R^2 \rho^2}{C_v} \qquad (19)$$

With a radius corresponding to the earth ($R = 6.370\,10^6$ m), corresponding medium density ($\rho = 5.523\,10^3$ kg/m$^3$) and average specific heat capacity ($C_v = 1.000$ J kg$^{-1}$ K$^{-1}$) is $\Delta T_{max} = 3.76\,10^4$ K (Hintz 2011) for an instantaneously grown body.

Celestial bodies of a size like the terrestrial planets and the Moon or bigger become certainly hot enough for entire melt. They can assume a hydrostatic equilibrium (nearly round) shape without have to overcome rigid-body forces.

On the other hand for a spherical celestial body with $R = 3\,10^5$ m and $\rho = 3{,}54\,10^3$ kg/m$^3$ the maximum temperature increase would be only $\Delta T_{max} = 114$ K and the cooling would be very fast. In this case melting needs further heat sources.

The cooling rate depends not only on the size and composition of the body, but also on the aggregate state and the temperature difference to the environment.

In the solid state there is a heat conduction regime. For a spherical solid body, the time to reach half the initial temperature difference results in relation to the transient average temperature from (CARSLOW & JAEGER 1959):

$$t_{1/2} \cong R^2 \frac{1}{\kappa \pi} \{\ln(6/\pi^2) - \ln 2\} \tag{20}$$

where $\kappa$ is the dissipation coefficient.

For a solid sphere of earth-like composition and density, the half-life period defined in this way is approximately 31 Ma (million years) at R = 100 km and already around 31 Ga (billion years) at R = 1000 km.

As soon as the temperature exceeds the solidus, the heat transport by convection competes with the heat conduction, with radii on the order of $10^3$ km even before the melt (sub-solidus convection). The surface heat flow is determined by the thickness of the thermal and mechanical boundary layer. In the case of slow convection, the mechanical boundary layer can be stable (stagnant lid regime, HTC-modelling (PROBSTHAIN 2013/2014)).

If the temperature increases further by accretion, the convection becomes more and more vigorous. The mechanical boundary layer becomes very short-lived, so that the surface temperature and with it the heat flux by radiation rapidly increase. Thus, the temperature of an Earth-like planet can be reduced to the present level within a period of some $10^8$ years, which is stabilized by the decay of long-lived radioactive isotopes.

### 2.4. Core mantle differentiation, mantle crust differentiation

The same problem as with the formation of a hydrostatic equilibrium shape arises with the differentiation of small celestial bodies into core and mantle, because the segregation requires at least a molten metallic phase. The solution is an extremely short accretion timescale of a few million years or less, which is commensurable with the half-lives of the short-lived elements (SLE) $^{26}$Al and $^{60}$Fe and the dissipation rate. Such an early core formation can be confirmed by Hf/W chronometry (WOOD. WALTER & WADE 2006, review).

The decay of radioactive isotopes is not only independent of pressure and temperature, but also of the size of the matrix. Therefore, even smaller objects can benefit from it. This makes this contribution to the thermal history of the growing celestial bodies particularly interesting for the present problem.

The isotopes of uranium, thorium and potassium, which are still active now, do not play a role here because they decay much too slowly. If at all, only short-lived isotopes such as $^{26}$Al, $^{60}$Fe, $^{53}$Mn, which are long extinguished, can have been effective.

The total amount of heat released during the decay is very high. From the data of their abundance, their half-life, and their initial heat flux in an earth-like matter (CZECHOWSKI &

WITEK 2015, Table 1) a maximum temperature increase of 73000 K results if the age of *Ca-Al-rich inclusions* CAI is used. However, the small aggregates initially formed dissipate the heat immediately to the environment so that a noticeable temperature increase only occurs with objects with radii of the order of magnitude of 1 km (MERK et al. 2002). In order to take this into account, numerical calculations usually start with a delay after CAI. However, there is considerably less heat available. After 1.8 Ma, for example, it is sufficient for a maximum temperature increase of 5700 K and after 20 Ma only of 10 K. The results of the model calculations therefore depend strongly on the choice of this parameter.

Further uncertainties exist with regard to the accretion law, that means the rate and the overall duration of the growth, as well as, not at least, to the ambient temperature, what is often not questioned.

Depending on the choice of the parameters, the asteroid 4 Vesta of about 560 km in diameter produces temperatures that at least lead to the partial melting of basaltic lava (NEUMANN et al. 2012), while in the Saturn satellite Mimas of almost 400 km in diameter they do not even melt the ice (CZECHOWSKI & WITEK 2015).

The resulting transient temperature distribution leads to the formation of a magma ocean beneath a solid crust (NEUMANN et al. 2012) , or a water mantle covered with a frozen crust (WAKITA & SEKIYA 2011). The metals in the magma ocean and the silicates in the water mantle can separate and sink forming a core.

But the crust is not mechanically stable. Continuous bombardment of the growing body by small planetesimals achieves local damage or melt of the crust. Underlying fluid can ascend to the surface and produce fresh crust (SAHIJPAL & BATHIA 2014), which is differentiated from the remaining mantle (MOSKOVITZ & GAIDOS 2011, GUPTA & SAHIJPAL 2010).

The heating by the decay of the SLE is mainly supported by conversion of the kinetic energy of impactors (CIESLA et al. 2013). Even long time after accretion when the shape of the body is stabilized local melting of the crust can occur by impact bombardment (ABRAMOV & MOJZSIS 2016). Impactors can penetrate the crust (COX et al. 2008).

## 2.5. Partial melt and hydrostatic equilibrium

Partial melt enables not only differentiation of small celestial bodies but also assuming of a hydrostatic equilibrium (nearly round) shape. As shown above even a solid crust is not an obstacle, because efficient resurfacing occurs. The crust could have been destabilized by mechanical damage and local melt. Besides impact bombardment vigorous convection of the melt beneath the crust, volcanism and stress by volume increase enhance the resurfacing.

The thickness and rigidity of the crust depend on the temperature of the environment, i. e. the temperature of the solar nebula. At very early stages of the thermal history the temperature may have been much higher than at present. CASSEN 2001 postulated the hypothesis that surviving planetary objects began to form as the nebula cooled from an early 'hot epoch'. In the initial state a dust condensation/evaporation front (1350 K isotherm) could extend over more than 3 AU.

Especially in the case of icy moons the ambient temperature is the crucial parameter in model calculations, determining whether ice can melt or not at moderate temperature increase. A span of 50 to 130 K is sometimes taken into account (WAKITA & SEKIYA 2011).

### 2.6. Some borderline cases

The recent NASA missions Dawn and Cassini brought detailed insights towards dimension, shape and surface structure of some asteroids within the belt between Mars and Jupiter and some moons of Saturn. The pictures received are very impressive and revealing. The largest asteroids are borderline cases for classification as dwarf planets (Table 2). Saturn's commensurable moons are references to distant and not-so-well-known objects like the TNOs.

Ceres' shape is accepted as nearly round. Therefore Ceres is the only one object in the asteroid belt, which is classified as dwarf planet. The oblate deviation from a perfect spherical shape is caused by spinning when a hydrostatic equilibrium of a liquid is assumed. Nevertheless the ratio of the surface area of such an oblate ellipsoid to the surface area of a sphere with an equal volume and mass is smaller than the lower limit value of a rocky or icy solid ($A_{pot}/A_{sph} < 1 + A_0/A_{sph}$). The shape must have been formed, while the object was at least partly melted. That is in accordance with the prediction that Ceres is highly differentiated with an icy crust, which has been undergoing steady resurfacing (CASTILLO-ROGEZ & MCCORD 2010, TRICARICO 2014, PARK et al. 2016). Regarding the assumption of its shape Ceres might behave more like an icy than a rocky body (NEVEU & DESCH 2016, FU 2015).

Pallas is the least explored of the three major asteroids (SCHMIDT et al. 2008). It shows a distinct deviation from a spherical shape and differs from Ceres in its size, density and composition.

Vesta of similar size as Pallas has been for a long time and is still now an object of special scientific interest. It is assumed as a model for the terrestrial planets and as the origin of the Vestoids in the asteroid belt and a class of meteorites (BURATTI et al. 2013). Although it was found that Vesta is differentiated in core, mantle and crust, it looks more like a potato than a sphere. Its present shape could be a remnant of a catastrophic collision when parts of the original already cooled and solidified body were split off. The surface of the northern 'hemisphere' seems to fit the surface of an oblate ellipsoid or even a much larger sphere (Fig. 3). But the volume of material split off would be orders of magnitude larger than that of the Vestoids known together (BURBINE et al. 2001). Two late heavy impacts have probably dug the southern pole to a depth of 80 km and caused widespread resurfacing with Veneneia and Rheasilvia formation (CLENET et al. 2014) .

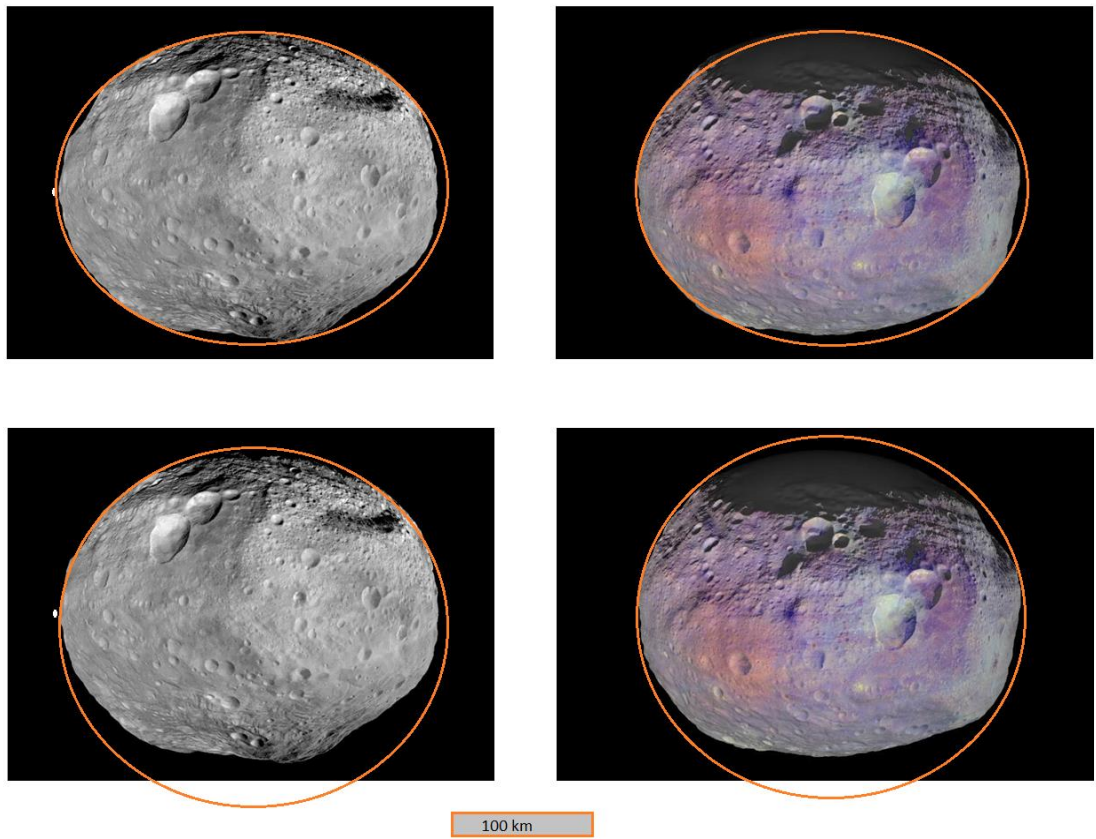

Fig. 3: The asteroid 4 Vesta. Is it a remnant of an originally 'nearly round' body? The 'northern hemisphere' fits an ellipsoid or even a sphere.

Enceladus and Mimas, although satellites are not candidates for dwarf planet, are interesting borderline cases with respect to the condition (b) of the planet definition. Enceladus is of the same size as Vesta, but has only half of its mass. Nevertheless its shape is concordant with a related fluid body in hydrostatic equilibrium. Enceladus is differentiated and up to now partly molten. Mimas, the next smaller satellite of Saturn is said to have never been melt in its history (Neumann et al. 2012, Czechowski & Witek 2015).. Its shape is dominated by the huge Herschel crater, which has a diameter one third the size of the whole satellite. Obviously a late heavy impact has altered an originally nearly spherical shape (Fig. 4). That alteration remains stable on a ground with a surface age of several billion years (Schmedemann et al. 2008). The original sphere could not have been formed while in solid state.

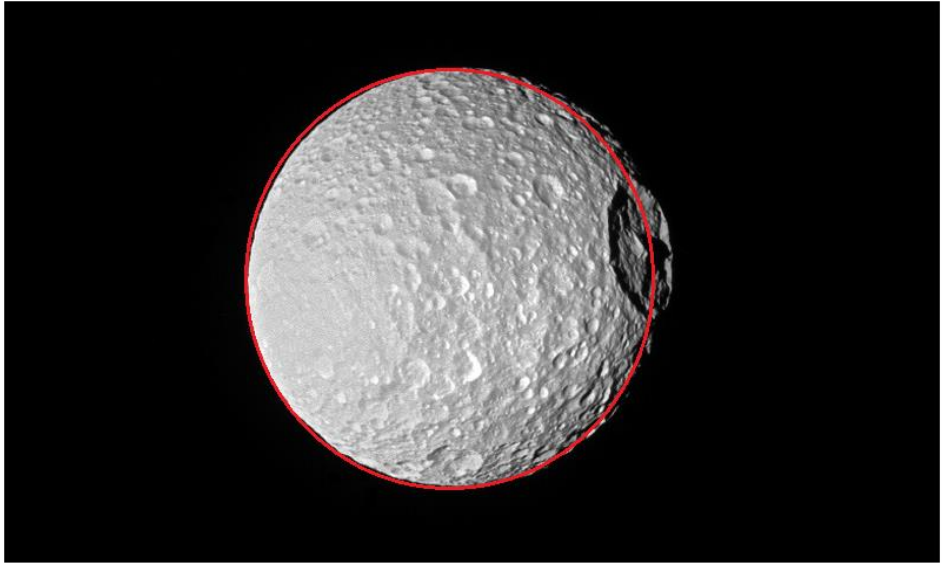

Fig. 4: The icy Saturn satellite Mimas. The huge Herschel crater could have altered the originally 'nearly round' shape of the body.

Like Mimas the Herschel crater the much larger Saturn moon *Iapetus supports a fossil bulge of over 30 km, and provides a benchmark for post global relaxation modification of the shape by impact cratering*, as THOMAS et al. 2007 stated.

The shapes and the other features of the TNO are not well known. But for the recently accepted dwarfs Makemake and Haumea deviations from the spherical shape could be established. Haumea is more flattened than all the other planets and dwarfs. It appears to be in hydrostatic equilibrium corresponding to a fast rotating liquid body, but certainly it is in the solid state.

Haumea seems to be unique in the solar system (BROWN 2011). All other planets and dwarfs are slightly flattened spheroids or spheres, while Haumea is a pronounced triaxial so-called Jacobi ellipsoid. It is questionable whether anyone has considered such an object as 'nearly round' in 2006, even though Haumea was discovered two years earlier.

### 3. Implications and conclusion

As shown above self-gravity can overcome rigid-body forces only while an overburden stress exists. At equilibrium the surface area of a solid is restricted, but its shape is variable. The hydrostatic equilibrium shape (ellipsoid) of a corresponding fluid can only be established at solid state, if the surface area of that ellipsoid is larger than the equilibrium surface area of the solid (Table 2, Fig. 2).

That would be applicable for the giants Jupiter and Saturn, if they would be completely in the solid state. But the rigid-body forces in their gaseous mantles are negligible. Therefore case (b2) for item (b) of the planet definition is only conditional applicable. The same is valid for the outer planets Uranus and Neptune.

Even the Earth has not mass enough for overcoming rigid-body forces in the estimated range of magnitude. But the Earth had been at least mostly fluid in its thermal history before solidification. That is case (b3).

The other planets, dwarfs and "nearly round" celestial bodies have also reached their original shape during their early history, even without overcoming rigid body forces. Late alterations of their shape remain stable, e. g. Herschel crater at the satellite Mimas. This is a sign that a hydrostatic equilibrium shape has been achieved before solidification.

Only the surface area of the fast rotating dwarf Haumea exceeds the limit for a solid body in equilibrium. Assuming that the fast rotation has started after the body was solidified, it has achieved a hydrostatic equilibrium shape of a corresponding fluid by overcoming the rigid-body forces. Apart from the fact that the form is anything but round, Haumea is the only example of (b1).

In summa none of the known celestial bodies fulfills the condition (b) of the planet definition in its stringent sense. Therefore we support the recommendation of SARMA et al. (2008) to delete this condition completely. In order to restrict the number of dwarfs a minimum mass equal to Ceres should be practicable as a new condition .

**Acknowledgment**

I thank Bernhard Steinberger for discussions and very useful hints.